\documentclass[review]{elsarticle}

\usepackage{lineno,hyperref}

\journal{Planetary and Space Science Journal}

\begin{document}

\begin{frontmatter}

\title{Probing the infrared counterparts of diffuse far-ultraviolet sources in the Galaxy}


\author[mainaddress]{Gautam Saikia\corref{mycorrespondingauthor}}
\cortext[mycorrespondingauthor]{Corresponding author}
\ead{gautamsaikia91@gmail.com}
\author[secondaryaddress]{P. Shalima}
\author[mainaddress]{Rupjyoti Gogoi}
\author[mainaddress]{Amit Pathak}

\address[mainaddress]{Department of Physics, Tezpur University, Napaam-784028, India}
\address[secondaryaddress]{Regional Institute of Education Mysore, Mysuru, Karnataka-570001, India}

\begin{abstract}

Recent availability of high quality infrared (IR) data for diffuse regions in the Galaxy and external galaxies have added to our understanding of interstellar dust. A comparison of ultraviolet (UV) and IR observations may be used to estimate absorption, scattering and thermal emission from interstellar dust. In this paper, we report IR and UV observations for selective diffuse sources in the Galaxy. Using archival mid-infrared (MIR) and far-infrared (FIR) observations from \textit{Spitzer Space Telescope}, we look for counterparts of diffuse far-ultraviolet (FUV) sources observed by the \textit{Voyager}, \textit{Far Ultraviolet Spectroscopic Explorer (FUSE)} and \textit{Galaxy Evolution Explorer (GALEX)} telescopes in the Galaxy. IR emission features at 8$\mu$m are generally attributed to Polycyclic Aromatic Hydrocarbon (PAH) molecules, while emission at 24$\mu$m are attributed to Very Small Grains (VSGs). The data presented here is unique and our study tries to establish a relation between various dust populations. By studying the FUV-IR correlations separately at low and high latitude locations, we have identified the grain component responsible for the diffuse FUV emission.

\end{abstract}

\begin{keyword}
\texttt{ISM: dust, infrared: ISM, Galaxy: extinction}
\end{keyword}

\end{frontmatter}


\section{Introduction}

The discovery of scattering and absorption of incident radiation, known collectively as extinction, provided the first definitive proof of the existence of interstellar dust (Trumpler 1930[1]). Our study of the properties of dust such as grain size, composition, etc., is mainly based on the spectroscopic absorption or emission features and thermal emission from the dust. There are many models that have been proposed to examine the dust in the diffuse interstellar medium (ISM) (Witt 2000[2]) which are mainly based on an analysis of extinction (Mathis et al. 1977[3], Hong \& Greenberg 1980[4], Draine \& Lee 1984[5], Duley et al. 1989[6], Kim et al. 1994[7], Mathis 1996[8], Li \& Greenberg 1997[9], Zubko 1999[10], Weingartner \& Draine 2001a[11]). The importance of interstellar dust can be gauged from the fact that it not only affects how we see our own Galaxy, but also affects the appearance of other galaxies by attenuating the short wavelength radiation from stars, and by re-emitting in the infrared (IR), far-infrared (FIR), sub-millimeter, millimeter, and microwave wavelength bands (Draine 2003[12]). Inspite of the existence of multiple models, the most accepted view is that the interstellar dust grains consist of amorphous silicates and some form of carbonaceous materials. \\

In the mid-infrared (MIR), we observe emission from small Polycyclic Aromatic Hydrocarbon (PAH) molecules (Allamandola et al. 1985[13]) and in the far-infrared (FIR) from solid grains with the grain size starting from a few tens of angstrom (Draine 2003[12]). These small grains are known as Very Small Grains (VSGs) and their emission is detected near 24$\mu$m. PAH molecules are electronically excited by the background UV photons. The excited PAHs emit in the MIR through infrared fluorescence. A significant amount of emission is found near 8$\mu$m (Wu et al. 2005[14]). \\

There have been large scale observations of diffuse FUV dust scattering in the Galaxy with the advent of space based telescopes such as the IR-based \textit{Spitzer Space Telescope} or the UV-based \textit{Voyager}, \textit{Far Ultraviolet Spectroscopic Explorer (FUSE)} and the more recent \textit{Galaxy Evolution Explorer (GALEX)} telescope. According to Bendo et al. (2008)[15], the stellar continuum-subtracted 8$\mu$m gives PAH emission, 24$\mu$m gives hot dust (VSG) emission and 160$\mu$m gives cold dust emission. In particular, the (8$\mu$m/24$\mu$m) surface brightness ratio is observed to be high in the diffuse ISM and low in bright star-forming regions. By studying the correlations between the MIR/FIR and FUV observations for the same locations in the Galaxy, we hope to probe the dust properties in the region and give accurate explanations for the observed correlation trends.\\

\section{Observations and Data Analysis}

We have looked for IR data corresponding to the FUV observations by \textit{Voyager} (Murthy et al. 1994[16], Murthy et al. 1999[17] and Sujatha et al. 2007[18]), \textit{FUSE} (Murthy \& Sahnow 2004[19] and Sujatha et al. 2007[18]) and \textit{GALEX} (Murthy 2014[20]) in our Galaxy. We found 48 MIR locations observed at 8$\mu$m and 80 MIR locations observed at 24$\mu$m in the \textit{Spitzer Heritage Archive (SHA)}. The 8$\mu$m observations have been taken using the \textit{Infrared Array Camera (IRAC)} which is on-board the \textit{Spitzer Space Telescope}. \textit{IRAC} (Fazio et al. 2004[21]) is a four-channel camera that provides simultaneous images at 3.6$\mu$m, 4.5$\mu$m, 5.8$\mu$m, and 8$\mu$m. The size of each of the four detector arrays in the camera is 256$\times$256 pixels, with a pixel size of 1.2$\times$1.2 arcsecs. The other 80 locations at 24$\mu$m have been observed by the \textit{Multiband Imaging Photometer for Spitzer (MIPS)}. The \textit{MIPS} (Rieke et al. 2004[22]) made observations and produced images and photometric data in three broad spectral bands in the MIR and FIR: 128$\times$128 pixels at 24$\mu$m with a pixel size of 2.45$\times$2.45 arcsec, 32$\times$32 pixels at 70$\mu$m with a pixel size of  4.0$\times$4.0 arcsec, and 2$\times$20 pixels at 160$\mu$m with a pixel size of 8.0$\times$8.0 arcsec. The FUV data and instruments have been well documented in the individual work (Murthy et al. 1994[16], Murthy et al. 1999[17], Murthy \& Sahnow 2004[19], Sujatha et al. 2007[18], Murthy 2014[20]) that have been used as source for this work. Briefly, the \textit{Voyager UVS} with a large field of view of 0.1$^{\circ}\times$0.87$^\circ$ observed diffuse radiation from 500-1600 \AA \hspace{0.1cm} with a resolution of about 38 \AA. It has long integration times resulting in a high sensitivity to diffuse radiation, viz. better than 100 photons cm$^{−-2}$ sr$^{-−1}$ s$^{−-1}$ \AA$^{−-1}$ (Sujatha et al. 2007[18]). The \textit{FUSE} telescope has the \textit{LWRS} (30''$\times$30'') aperture on-board and the four \textit{FUSE} spectrographs having a resolution ($\lambda/\Delta\lambda$) of about 20000 cover the wavelength range 850-1167 \AA. It can detect background levels of 2000 photons cm$^{−-2}$ sr$^{-−1}$ s$^{−-1}$ \AA$^{−-1}$ (Sujatha et al. 2007[18]). Murthy \& Sahnow 2004[19] have described the method of extraction of diffuse surface brightness from FUSE spectra by analysis of the background observations. The spectra is then collapsed into two wavelength bands per detector by treating the FUSE spectrum as a broadband photometric observation and excluding the terrestrial air glow lines (primarily Ly$\beta$). The \textit{GALEX} telescope instrument having spatial resolution of 5''-–10'' has two detectors: FUV at 1344–-1786 \AA \hspace{0.1cm} and NUV at 1771-–2831 \AA, which image a 0.6$^\circ$ radius field. It detects a diffuse signal of 100 photons cm$^{−-2}$ sr$^{-−1}$ s$^{−-1}$ \AA$^{−-1}$  in a typical \textit{AIS} observation  (Murthy 2014[20]).\\

We have used all the 48 locations with 8$\mu$m data and all the 80 locations with 24$\mu$m data. We have calculated the flux at these locations using aperture photometry technique and then converted them to intensities. The images have not been convolved. We have taken post basic calibrated data (pbcd) and treated the images at different wavelengths independently. After computing the IR intensities, we have calculated the Spearman's rank correlation coefficient ($\rho$) which is a non-parametric version of rank correlation. We have calculated the correlations among intensities and since the spatial resolution of all the telescopes are comparable, this should not affect the results. The Spearman's rank correlation is a simple and reliable method of testing both the strength and direction (positive or negative) of the monotonic relationship between two variables rather than the linear relationship between them. It does not assume any model, like a straight line fit, and hence it is non-parametric. We have used Spearman's rank correlation because monotonicity is `less restrictive' than that of a linear relationship, i.e. we might get a pattern among our observed data that is monotonic, but not linear, and so it still tells us that they are related (Bevington 2003[23]). The Spearman's rank correlation coefficient is calculated using the following relation:
$$ \rho = 1 - \frac{6 \Sigma d^2}{n^3  - n}$$
where, $\Sigma$ = sum, d = difference between two ranks, n = no. of pairs of data.\\

The Spearman correlation coefficient, $\rho$, can take values inside the interval [-1, 1]. A value of +1 indicates a perfect association of ranks (the value of one variable increases with the other), zero indicates that there is no connection between ranks and -1 indicates a perfect negative association of ranks (the value of one variable increases with decrease in value of the other). As the value of $\rho$ gets closer to zero, the association between the ranks becomes weaker. Once we have calculated the rank correlation, we must test to see how likely it is that our calculation is not just the result of chance which is called significance testing. It considers our result in relation to how much data we have. The probability ($\delta$) is a result of this significance testing which tells us the standard error in the calculation of the rank correlation coefficient. It is calculated as:
$$ \delta = \frac{0.6325}{\sqrt{(n-1)}} $$
A lower value of probability ($\delta$) indicates graeter reliability in the observed value of rank correlation coefficient. \\

\begin{figure}[h!]
\centering
\includegraphics[width=12cm,height=5cm]{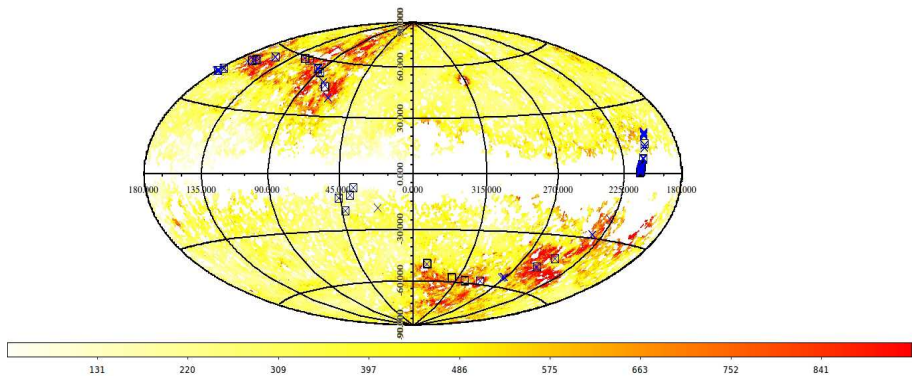}
\caption{\normalsize Our locations plotted on an Aitoff with Galactic coordinates. The 8$\mu$m locations are represented as squares and 24$\mu$m locations as crosses. This is an all-sky map of the FUV diffuse sky from Murthy 2014[20].}
\label{aitoff}
\end{figure}

Our locations include the \textit{Coalsack} nebula, which is one of the prominent dark nebulae in the southern sky of the \textit{Milky Way}. The emission from the nebula is mainly due to the forward scattering of light from the bright stars behind it as mentioned by Murthy et al. 1994[16]. While \textit{Infrared Spectrograph (IRS)} from \textit{Spitzer} provides the most accurate way to study the dust features, it would be useful to use some combination of \textit{IRAC} and \textit{MIPS} images since having a photometric probe would allow for a larger range of objects and environments to be studied given that imaging is faster to acquire and can probe regions that are too faint for spectroscopy. The 8$\mu$m band of the \textit{IRAC} is designed to cover the 7.7$\mu$m PAH feature, while the 24$\mu$m of \textit{MIPS} just covers the dust continuum from VSGs for local galaxies, avoiding the contamination of most of the MIR emission or absorption lines. Figure \ref{aitoff} shows our locations plotted on an all-sky map of the FUV diffuse sky from Murthy 2014[20].

\section{Results and Discussion}

We have first obtained the intensity values for the data that we have observed. Next, we have calculated the rank correlation between our FUV and MIR data. Since the source of FUV radiation could be different at low and high latitudes with dust grains being the main contributor at moderate to low latitudes, we have separated our data into high and low Galactic latitude locations and calculated the correlations separately. Locations within the range $-20^\circ \leq|b|\leq 20^\circ$ are considered to be low latitude and those in the range $|b|>20^\circ$ to $|b|<-20^\circ$ are considered to be high latitude, `b' being the Galactic latitude. We have also included the correlations between the hydrogen column density, \textit{N(H)} and the MIR intensities. In order to calculate the value of \textit{N(H)}, we have used the all-sky 100$\mu$m dust emission map by Schlegel et al. (1998)[24] to extract the color excess, E(B-V) for our locations. We have then used the following relation for total to selective extinction at the B and V filter bands in order to obtain the total extinction A(V) for each location:
$$\frac{A(V)}{E(B-V)} = R(V)$$
where R(V) = 3.1 for the \textit{Milky Way} (Draine 2003[12]).
The value of A(V) is then substituted in the following:
$$\frac{\textit{N(H)}}{A(V)} = 1.8 \times 10^{21} \hspace{0.2cm} atoms \hspace{0.1cm} cm^{-2} \hspace{0.1cm} mag^{-1}$$
to obtain the required value of \textit{N(H)} for that particular location.\\

The optical depth is given by $\tau (\lambda) = \textit{N(H)} \times \sigma$, where $\lambda = 1500$\AA \hspace{0.1cm} and $\sigma$ is the extinction cross-section. Here, as the number density of particles increases, the scattering also increases ($n\sigma$). Now, if $\sigma$ goes higher, there will be multiple scattering that dominates, much before $\tau$ approaches 1 since the albedo will be higher. At high values of $\sigma$ and resultant $\tau$, we will have significantly high extinction ($e^{-\tau}$). But multiple scattering will dominate at low latitudes and lead to more absorption. Hence, the scattered intensities are no longer linear with $n$, i.e. $\tau$ and this will heavily affect the correlation values. It will be linear only in case of single scattering. At low values, $\tau\ll$1, single scattering dominates. For the lower latitude locations with a cutoff value of $\textit{N(H)}=9\times 10^{21}$ cm$^{-2}$, we have $\tau = 0.0916$, while for the higher latitude locations with a cutoff value of $\textit{N(H)}=3\times 10^{20}$ cm$^{-2}$, we have $\tau = 0.00305$ which correspond to the values of $\sigma$ at $\lambda = 1500$\AA \hspace{0.1cm} (Draine 2003[12]). Hence, we use a cutoff (k) on \textit{N(H)} to keep multiple scattering in check by keeping $\tau$ low such that it does not affect our observed correlations.\\ 

\begin{figure}[h]
\centering
\includegraphics[width=5cm,height=5cm]{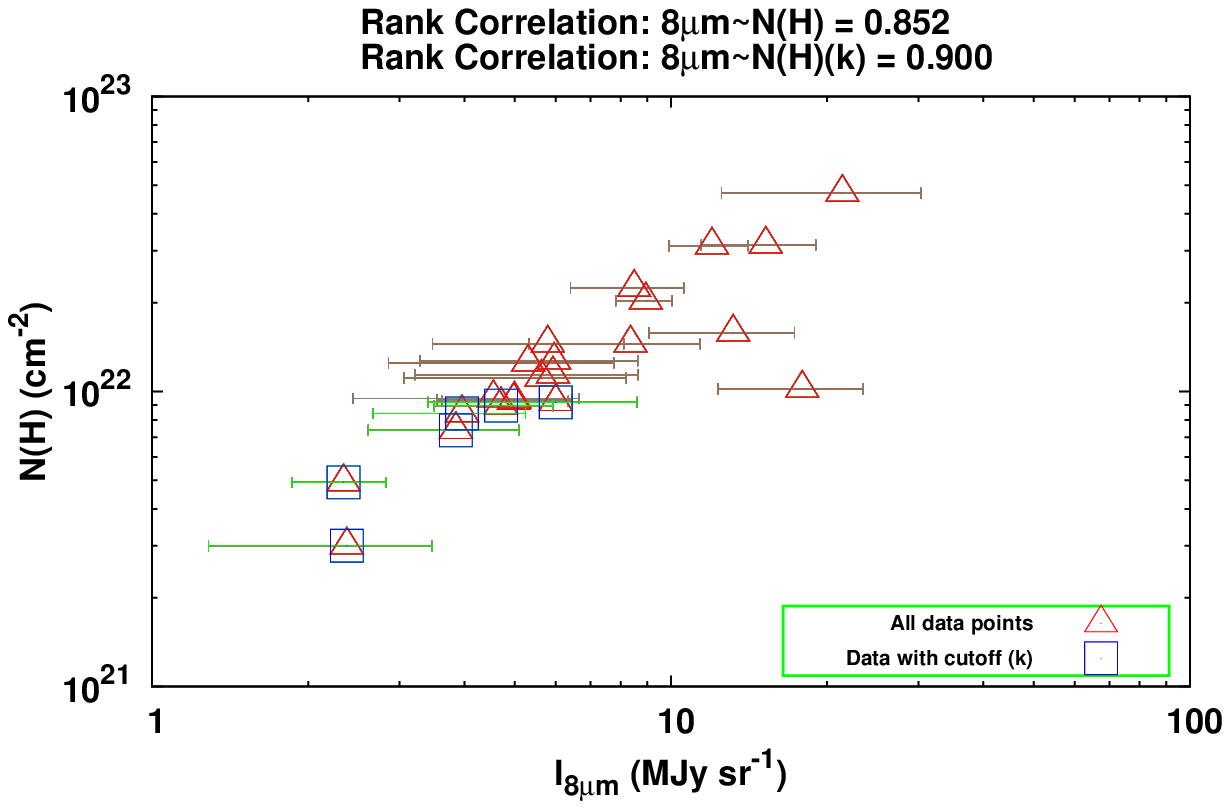}
\includegraphics[width=5cm,height=5cm]{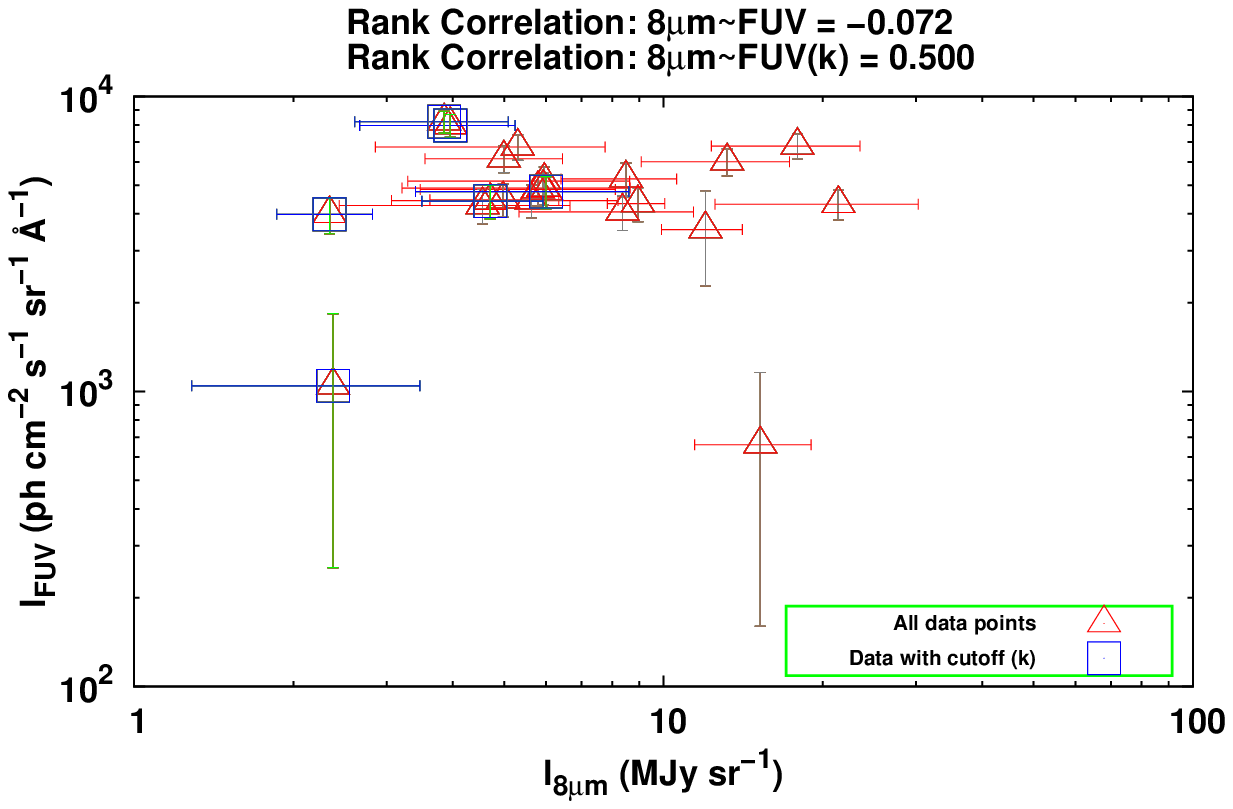}
\caption{\normalsize Correlations plotted for 8$\mu$m intensity vs \textit{N(H)} and FUV intensity at low latitude. Here k means a cutoff for \textit{N(H)} at $9\times 10^{21}$ cm$^{-2}$.}
\label{only8fig_lower}
\end{figure}

\begin{figure}[h]
\centering
\includegraphics[width=5cm,height=4.36cm]{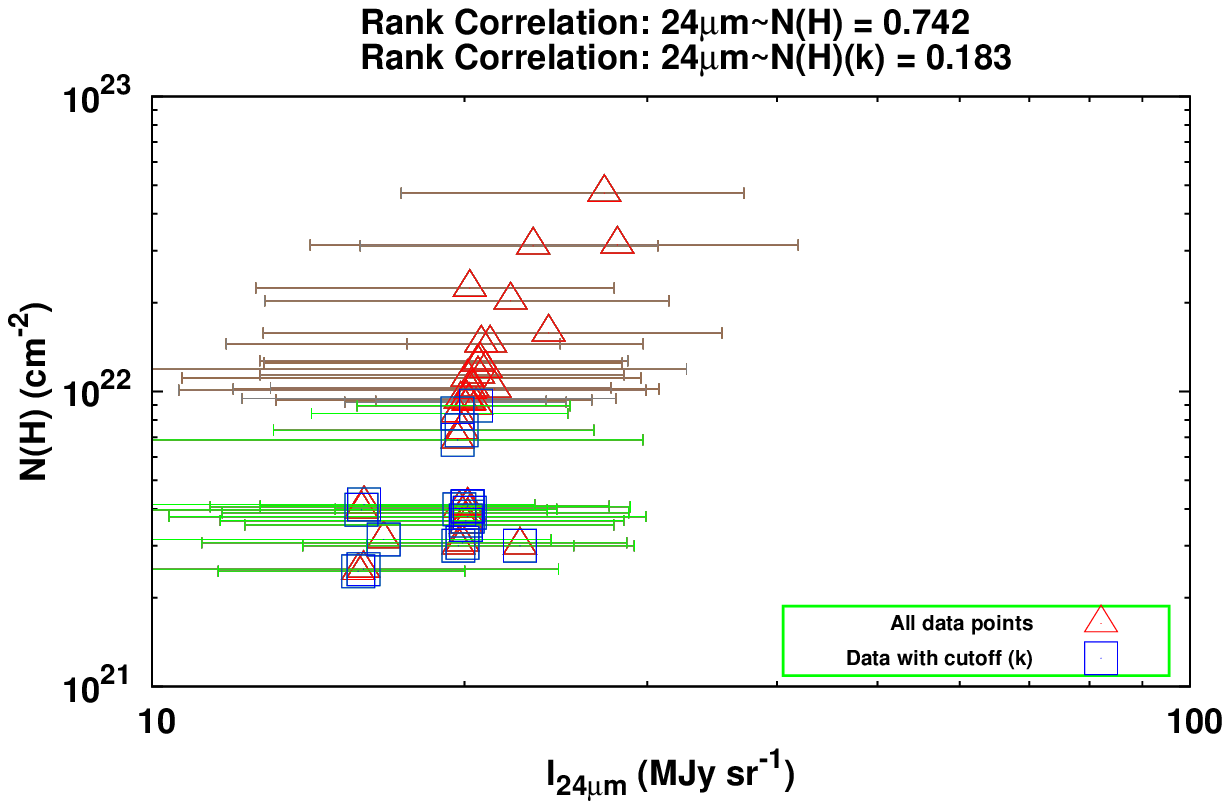}
\includegraphics[width=5cm,height=4.36cm]{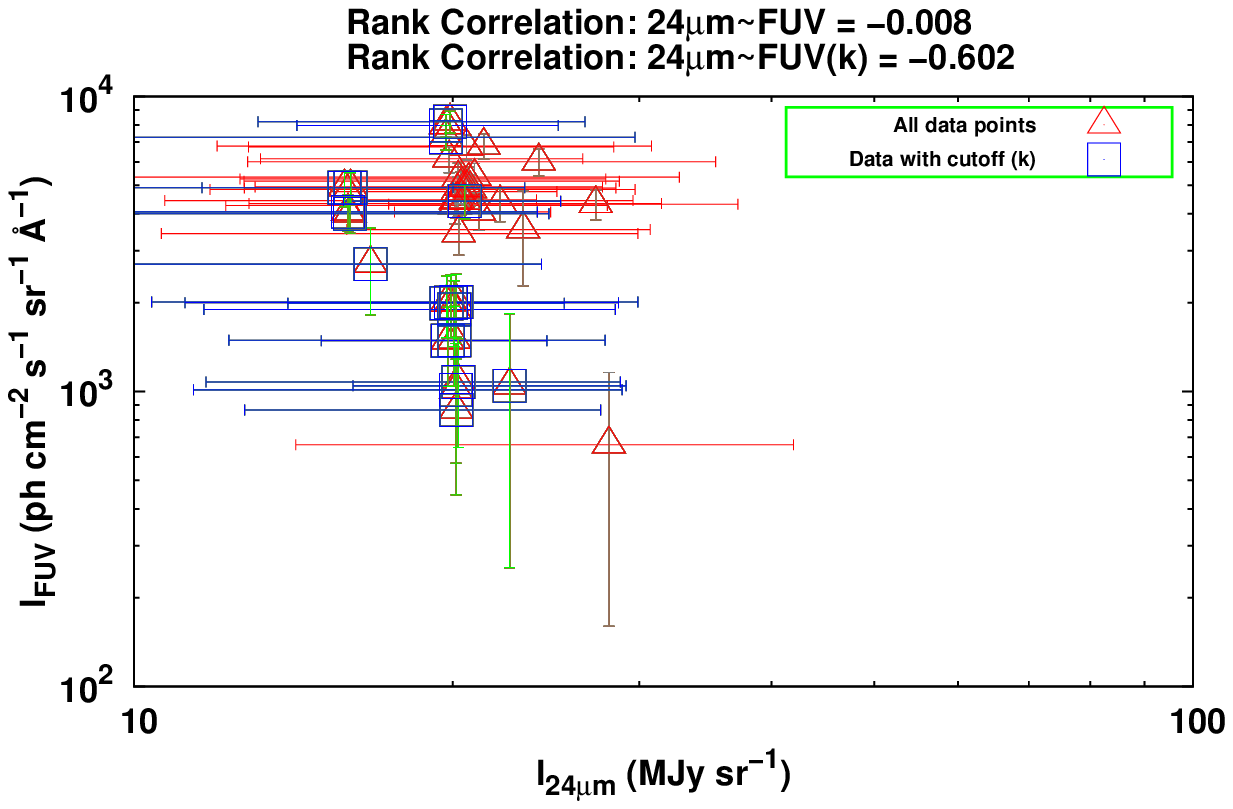}
\caption{\normalsize Correlations plotted for 24$\mu$m intensity vs \textit{N(H)} and FUV intensity at low latitude. Here k means a cutoff for \textit{N(H)} at $9\times 10^{21}$ cm$^{-2}$.}
\label{only24fig_lower}
\end{figure}

\begin{figure}[h]
\centering
\includegraphics[width=5cm,height=5cm]{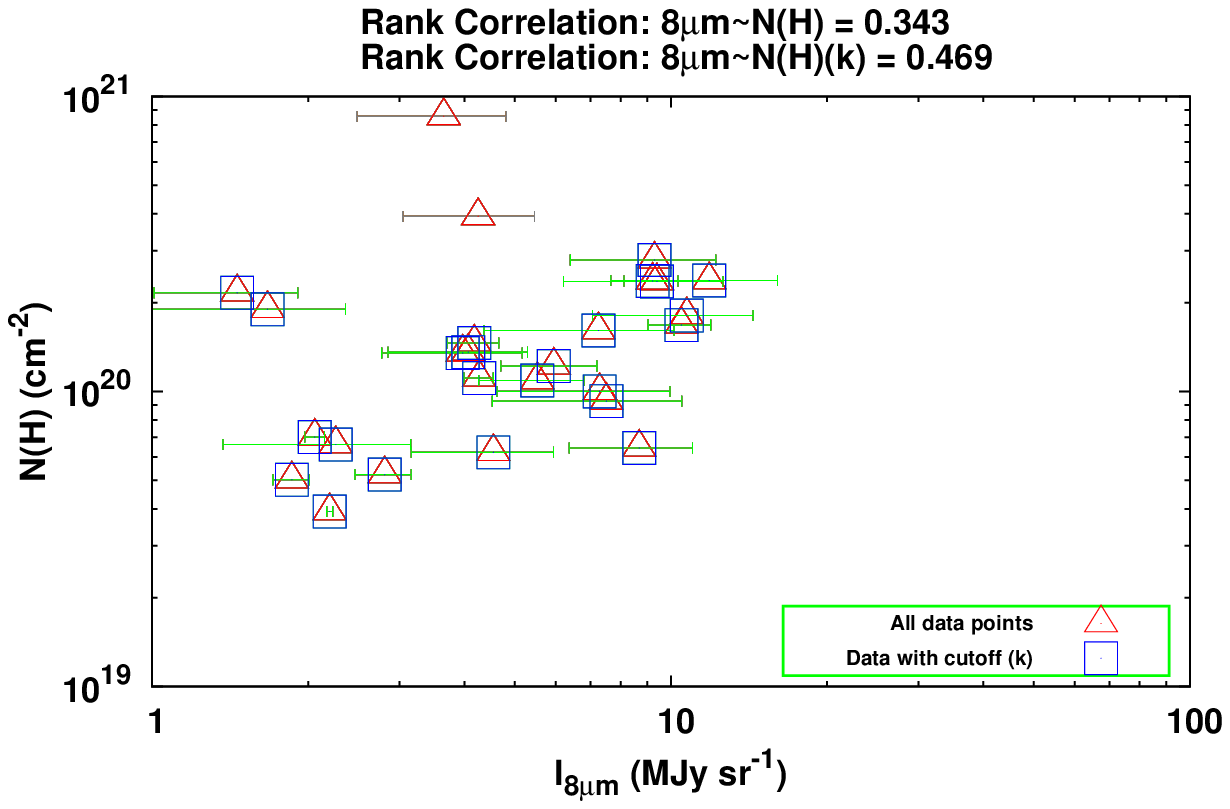}
\includegraphics[width=5cm,height=5cm]{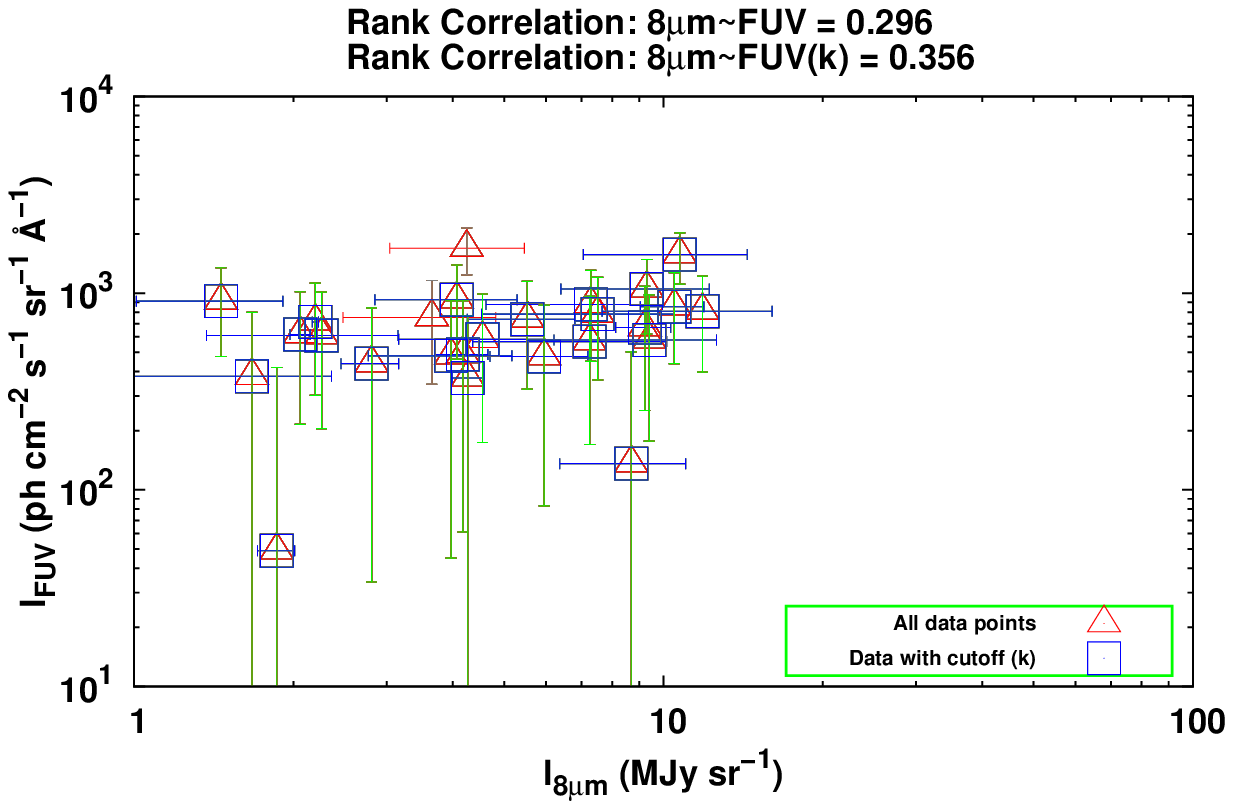}
\caption{\normalsize Correlations plotted for 8$\mu$m intensity vs \textit{N(H)} and FUV intensity at high latitude. Here k means a cutoff for \textit{N(H)} at $3\times 10^{20}$ cm$^{-2}$.}
\label{only8fig_higher}
\end{figure}

\begin{figure}[h]
\centering
\includegraphics[width=5cm,height=5cm]{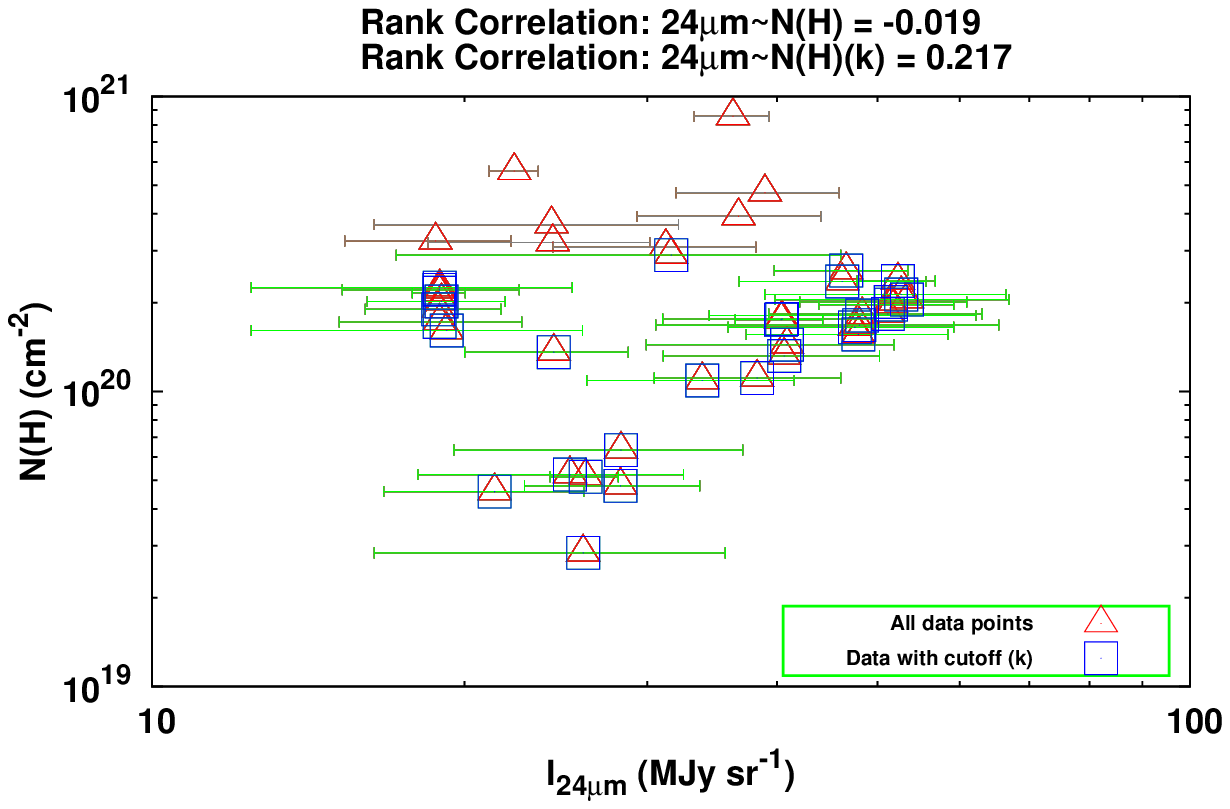}
\includegraphics[width=5cm,height=5cm]{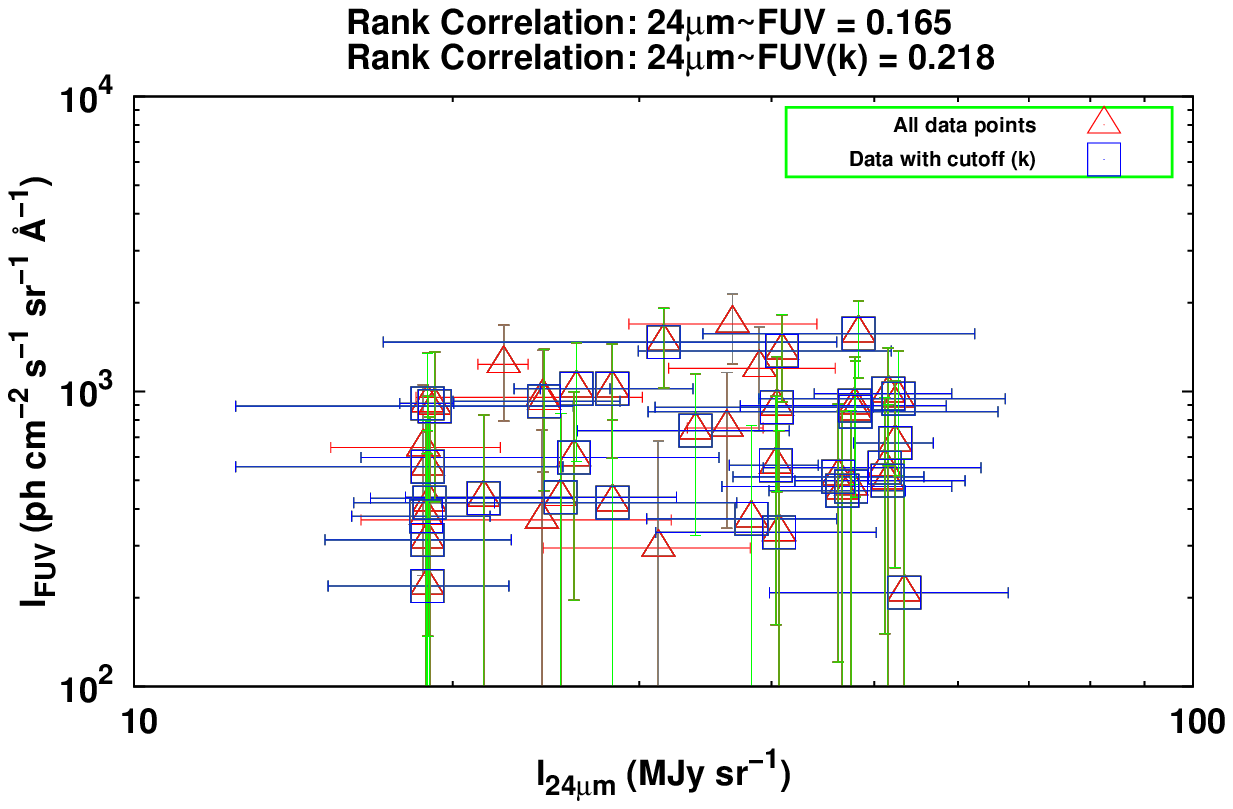}
\caption{\normalsize Correlations plotted for 24$\mu$m intensity vs \textit{N(H)} and FUV intensity at high latitude. Here k means a cutoff for \textit{N(H)} at $3\times 10^{20}$ cm$^{-2}$.}
\label{only24fig_higher}
\end{figure}

We have 48 locations having data for both 8$\mu$m and FUV. Among these, 22 are low latitude locations while 26 are high latitude locations. We have calculated their Spearman's rank correlations with and without a cutoff (k) on the value of \textit{N(H)} as listed in Table \ref{only_lower} and Table \ref{only_higher}. We have 80 locations having 24$\mu$m data among which 39 are low latitude locations while 41 are high latitude locations. The low latitude correlations are listed in Table \ref{only_lower} and the plots are shown in Figure \ref{only8fig_lower} and Figure \ref{only24fig_lower}. The axes labeled as `I' with a subscript indicate the intensity at that particular wavelength, e.g. I$_{8\mu m}$ indicates intensity at 8$\mu$m. The high latitude correlations are listed in Table \ref{only_higher} and shown in Figure \ref{only8fig_higher} and Figure \ref{only24fig_higher}.\\ 

\begin{center}
\begin{table}[h]
\centering
\begin{tabular}{ccc}
\hline
IR-IR(UV) & Rank correlation & Probability\\
\hline
$8.0 \sim \textit{N(H)}$&0.852&4.776e-07\\
$8.0 \sim \textit{N(H)}(k)$&0.900&0.0374\\
$8.0 \sim FUV$&-0.072&0.751\\
$8.0 \sim FUV(k)$&0.500&0.391\\
\hline
$24.0 \sim \textit{N(H)}$&0.742&6.617e-08\\
$24.0 \sim \textit{N(H)}(k)$&0.183&0.454\\
$24.0 \sim FUV$&-0.008&0.963\\
$24.0 \sim FUV(k)$&-0.602&0.006\\
\hline
\end{tabular}
\caption{\label{only_lower} \normalsize Rank correlations for the 22 low latitude locations having 8$\mu$m data and 39 low latitude locations having 24$\mu$m data. Here k means a cutoff for \textit{N(H)} at $9\times 10^{21}$ cm$^{-2}$.}
\end{table}
\end{center}

From our low latitude correlation results (Table \ref{only_lower}), we see that the 8$\mu$m intensity is better correlated to the \textit{N(H)}. The \textit{N(H)} is a direct measure of the intensity at 100$\mu$m. Hence, the 8$\mu$m is very well correlated to the 100$\mu$m intensity. This is in-line with the existing result that PAHs at MIR 8$\mu$m are associated with cold diffuse ISM observed at FIR 100$\mu$m (Seon et al. 2011a[25], Seon et al. 2011b[26], Hamden et al. 2013[27]). The comparatively lower correlation between 24$\mu$m and 100$\mu$m is also expected because these are associated with grains in different environments, i.e. the 24$\mu$m emission is from hot VSG populations (Calzetti et al. 2007[28], Prescott et al. 2007[29]) while the 100$\mu$m emission is from larger/colder grains. The FUV scattering is negatively correlated to the MIR 8$\mu$m and 24$\mu$m emission, i.e. as FUV emission increases, the MIR emission decreases and vice versa. This could be because of absorption which reduces the FUV intensity in regions of high dust density. This may also be due to the fact that PAHs are prone to destruction in the presence of high UV radiation fields (Madden 2000[30], Galliano et al. 2005[31]). In addition to this, we see that the correlation values seem to increase or get better for the 8$\mu$m data when using a cutoff (k) on the value of \textit{N(H)}. Similarly we see an improvement in the negative correlation between 24$\mu$m and FUV emission for regions of low \textit{N(H)} at low latitudes. Note that these are also the regions where there is no correlation between the \textit{N(H)} and 24$\mu$m intensities. This clearly shows that \textit{N(H)} which is calculated from the 100$\mu$m emission due to colder dust grains is unrelated to the FUV intensities in regions where there is 24$\mu$m emission. Another thing to note here is the probability value of the correlations obtained which are very low for the 8$\mu$m--\textit{N(H)} and 24$\mu$m--\textit{N(H)} and thus supporting the observed correlation values. The probability values are also low for the 8$\mu$m--\textit{N(H)}(k) and 24$\mu$m--FUV(k), thus asserting the significance of the observed correlations. The probability is however high for the 24$\mu$m--FUV (without cutoff) which tells us that the observed correlation value has a reduced credibility. Thus, we can say that both 8$\mu$m and 24$\mu$m have good positive correlations with \textit{N(H)} (with the 8$\mu$m correlation being comparatively better), while 24$\mu$m has a good negative correlation with the FUV data.\\

\begin{center}
\begin{table}[h]
\centering
\begin{tabular}{ccc}
\hline
IR-IR(UV) & Rank correlation & Probability\\
\hline
$8.0 \sim \textit{N(H)}$&0.343&0.086\\
$8.0 \sim \textit{N(H)}(k)$&0.469&0.021\\
$8.0 \sim FUV$&0.296&0.141\\
$8.0 \sim FUV(k)$&0.356&0.087\\
\hline
$24.0 \sim \textit{N(H)}$&-0.019&0.906\\
$24.0 \sim \textit{N(H)}(k)$&0.217&0.224\\
$24.0 \sim FUV$&0.165&0.301\\
$24.0 \sim FUV(k)$&0.218&0.223\\
\hline
\end{tabular}
\caption{\label{only_higher} \normalsize Rank correlations for the 28 high latitude locations having 8$\mu$m data and 41 high latitude locations having 24$\mu$m data. Here k means a cutoff for \textit{N(H)} at $3\times 10^{20}$ cm$^{-2}$.}
\end{table}
\end{center}

On the other hand, from our correlation results at high latitude locations (Table \ref{only_higher}), we do not get any significant correlation for either 8$\mu$m or 24$\mu$m data. The probability values are also low enough in almost all cases meaning that the observed correlation values are true. As we move to higher latitudes, the correlations go on getting lower probably due to decreasing abundance of interstellar PAHs and VSGs. This supports the presence of interstellar dust in the lower latitude locations. At high latitutdes we do not expect much dust and/or PAHs and we expect much less FUV radiation as much less population of young stars is expected compared to low latitudes. The FUV radiation observed at high latitudes may be the scattered light from far-off dust (Witt et al. 2010[32], Seon et al. 2011b[26]).\\

\begin{figure}[h!]
\centering
\includegraphics[width=10cm,height=6cm]{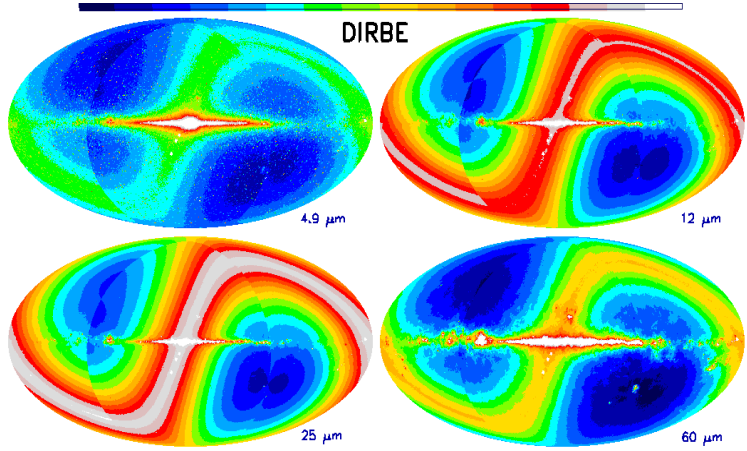}
\caption{\normalsize DIRBE maps (Sano et al. 2015[34]) explaining our observed correlations.}
\label{DIRBE}
\end{figure}

 Murthy (2014[20], 2016[33]) has observed and modelled (using Monte Carlo models with multiple scattering) the FUV and NUV all-sky GALEX data in our Galaxy and identified high latitude regions where there is an offset between the model predictions and the observed data (red colored patches in Figure \ref{aitoff}). They seem to coincide with regions that are not in the ecliptic plane, i.e. plane of the solar system where planetary dust contributes. We see this in NASA's Diffuse Infrared Background Experiment (DIRBE) maps as shown in Figure \ref{DIRBE} (Sano et al. 2016[34]). The bright diagonal line is the ecliptic. The regions in the other two quadrants do not have contribution from the ecliptic plane and so they truly represent pure interstellar dust with some extragalactic contribution. Since some of our diffuse source locations fall into these regions of asymmetry, this might explain the low correlations we are getting at such high latitudes.\\

\newpage

\section{Conclusions}

The important conclusions of this work are:

\begin{itemize}
\item The 8$\mu$m intensity shows higher correlation with the \textit{N(H)} or 100$\mu$m intensity (Fig. \ref{only8fig_lower}) as compared to 24$\mu$m with \textit{N(H)} (Fig. \ref{only24fig_lower}) which supports previous theory that PAH emission at 8$\mu$m band shows good correlation with cold dust emission and VSG emission at 24$\mu$m is hot dust, i.e. both belong to different dust populations.

\item The FUV scattering is negatively correlated to both the MIR 8$\mu$m (Fig. \ref{only8fig_lower}) and MIR 24$\mu$m (Fig. \ref{only24fig_lower}) emission which implies that the FUV gets absorbed and then re-emitted as IR by the dust and that FUV emission is complementary to the 8$\mu$m and 24$\mu$m emission. The negative correlation is better in case of FUV and 24$\mu$m which shows that the FUV absorption is predominantly due to hot dust grains. This is also evident in the lack of correlation between the \textit{N(H)} and 24$\mu$m emission for locations where the FUV--24$\mu$m correlation is significant.

\item The correlations are significantly better in low latitude locations (Fig. \ref{only8fig_lower}, \ref{only24fig_lower}) as compared to high latitude locations (Fig. \ref{only8fig_higher}, \ref{only24fig_higher}) for both 8$\mu$m and 24$\mu$m which indicates decreasing abundance of interstellar PAHs and VSGs at high latitudes. Earlier work done by Murthy (2016[33]) and others have not found any correlation among FUV and 100$\mu$m which could be attributed to two reasons: high latitude and VSG dominance since we did find a correlation with 24$\mu$m emitting grains (Table \ref{only_lower}). 

\item The DIRBE maps (Fig. \ref{DIRBE}) seem to indicate a remote origin for the observed FUV intensities at high galactic latitudes and hence support the claim for the extragalactic contribution at locations which are far from the ecliptic plane.
\end{itemize}

\section*{Acknowledgements}

This work is based in part on observations made with the \textit{Spitzer Space Telescope}, which is operated by the Jet Propulsion Laboratory, California Institute of Technology under a contract with NASA. This research has made use of the \textit{NED} and the \textit{SIMBAD} databases. PS would like to thank DST Fast Track (grant no SR/FTP/PS-68/2008) programme for financial support. RG and AP would like to thank the IUCAA Visitors' programme for their support and hospitality.\\

\section*{References}

\begin{enumerate}

\item R. J. Trumpler, Spectrophotometric Measures of Interstellar Light Absorption, 42 (1930) 267. doi:10.1086/124051.

\item A. N. Witt, Overview of Grain Models, in: Y. C. Minh, E. F. van Dishoeck (Eds.), From Molecular Clouds to Planetary, Vol. 197 of IAU Symposium, 2000, p. 317. arXiv:astro-ph/9910116.

\item J. S. Mathis, W. Rumpl, K. H. Nordsieck, The size distribution of interstellar grains, 217 (1977) 425–433. doi:10.1086/155591.

\item S. S. Hong, J. M. Greenberg, A unified model of interstellar grains - A connection between alignment efficiency, grain model size, and cosmic abundance, 88 (1980) 194–202.

\item B. T. Draine, H. M. Lee, Optical properties of interstellar graphite and silicate grains, 285 (1984) 89–108. doi:10.1086/162480.

\item W. W. Duley, A. P. Jones, D. A. Williams, Hydrogenated amorphous carbon-coated silicate particles as a source of interstellar extinction, 236 (1989) 709–725.

\item S. H. Kim, P. G. Martin, The size distribution of interstellar dust particles as determined from polarization: Infinite cylinders, 431 (1994) 783–796. doi:10.1086/174529.

\item J. S. Mathis, Dust Models with Tight Abundance Constraints, 472 (1996) 643. doi:10.1086/178094.

\item A. Li, J. M. Greenberg, A unified model of interstellar dust., 323 (1997) 566–584.

\item V. G. Zubko, On the Model of Dust in the Small Magellanic Cloud, 513 (1999) L29–L32. arXiv:astro-ph/9812460, doi:10.1086/311903.

\item J. C. Weingartner, B. T. Draine, Dust Grain-Size Distributions and Extinction in the Milky Way, Large Magellanic Cloud, and Small Magellanic Cloud, 548 (2001) 296–309. arXiv:astro-ph/0008146, doi:10.1086/318651.

\item B. T. Draine, Interstellar Dust Grains, 41 (2003) 241–289. arXiv:astro-ph/0304489, doi:10.1146/annurev.astro.41.011802.094840.

\item L. J. Allamandola, A. G. G. M. Tielens, J. R. Barker, Polycyclic aromatic hydrocarbons and the unidentified infrared emission bands - Auto exhaust along the Milky Way, 290 (1985) L25–L28. doi:10.1086/184435.

\item J. Wu, N. J. Evans, II, Y. Gao, P. M. Solomon, Y. L. Shirley, P. A. Vanden Bout, Connecting Dense Gas Tracers of Star Formation in our Galaxy to High-z Star Formation, 635 (2005) L173–L176. arXiv:astro-ph/0511424, doi:10.1086/499623.

\item G. J. Bendo, B. T. Draine, C. W. Engelbracht, G. Helou, M. D. Thornley, C. Bot, B. A. Buckalew, D. Calzetti, D. A. Dale, D. J. Hollenbach, A. Li, J. Moustakas, The relations among 8, 24 and 160 μm dust emission within nearby spiral galaxies, 389 (2008) 629–650. arXiv:0806.2758, doi:10.1111/j.1365-2966.2008.13567.x.

\item J. Murthy, R. C. Henry, J. B. Holberg, Voyager observations of dust scattering near the Coalsack nebula, 428 (1994) 233–236. doi:10.1086/174234.

\item J. Murthy, D. Hall, M. Earl, R. C. Henry, J. B. Holberg, An Analysis of 17 Years of Voyager Observations of the Diffuse Far-Ultraviolet Radiation Field, 522 (1999) 904–914. doi:10.1086/307652.

\item N. V. Sujatha, J. Murthy, P. Shalima, R. C. Henry, Measurement of Dust Optical Properties in the Coalsack Nebula, 665 (2007) 363–368. arXiv:0705.1752, doi:10.1086/519439.

\item J. Murthy, D. J. Sahnow, Observations of the Diffuse Far-Ultraviolet Background with the Far Ultraviolet Spectroscopic Explorer, 615 (2004) 315–322. arXiv:astro-ph/0407519, doi:10.1086/424441.

\item J. Murthy, GALEX Diffuse Observations of the Sky: The Data, 213 (2014) 32. arXiv:1406.5680, doi:10.1088/0067-0049/213/2/32.

\item G. G. Fazio, J. L. Hora, L. E. Allen, M. L. N. Ashby, P. Barmby, L. K. Deutsch, J.-S. Huang, S. Kleiner, M. Marengo, S. T. Megeath, G. J. Melnick, M. A. Pahre, B. M. Patten, J. Polizotti, H. A. Smith, R. S. Taylor,
Z. Wang, S. P. Willner, W. F. Hoffmann, J. L. Pipher, W. J. Forrest, C. W. McMurty, C. R. McCreight, M. E. McKelvey, R. E. McMurray, D. G. Koch, S. H. Moseley, R. G. Arendt, J. E. Mentzell, C. T. Marx, P. Losch, P. Mayman, W. Eichhorn, D. Krebs, M. Jhabvala, D. Y. Gezari, D. J. Fixsen, J. Flores, K. Shakoorzadeh, R. Jungo, C. Hakun, L. Workman, G. Karpati, R. Kichak, R. Whitley, S. Mann, E. V. Tollestrup, P. Eisenhardt, D. Stern, V. Gorjian, B. Bhattacharya, S. Carey, B. O. Nelson, W. J. Glaccum, M. Lacy, P. J. Lowrance, S. Laine, W. T. Reach, J. A. Stauffer, J. A. Surace, G. Wilson, E. L. Wright, A. Hoffman, G. Domingo, M. Cohen, The Infrared Array Camera (IRAC) for the Spitzer Space Telescope, 154 (2004) 10–17. arXiv:astro-ph/0405616, doi:10.1086/422843.

\item G. H. Rieke, E. T. Young, C. W. Engelbracht, D. M. Kelly, F. J. Low, E. E. Haller, J. W. Beeman, K. D. Gordon, J. A. Stansberry, K. A. Misselt, J. Cadien, J. E. Morrison, G. Rivlis, W. B. Latter, A. Noriega-Crespo, D. L. Padgett, K. R. Stapelfeldt, D. C. Hines, E. Egami, J. Muzerolle, A. Alonso-Herrero, M. Blaylock, H. Dole, J. L. Hinz, E. Le Floc’h, C. Papovich, P. G. P´erez-Gonz´alez, P. S. Smith, K. Y. L. Su, L. Bennett, D. T. Frayer, D. Henderson, N. Lu, F. Masci, M. Pesenson, L. Rebull, J. Rho,J. Keene, S. Stolovy, S. Wachter, W. Wheaton, M. W. Werner, P. L. Richards, The Multiband Imaging Photometer for Spitzer (MIPS), 154 (2004) 25–29. doi:10.1086/422717.

\item P. R. Bevington, D. K. Robinson, Data reduction and error analysis for the physical sciences, 2003.

\item D. J. Schlegel, D. P. Finkbeiner, M. Davis, Maps of Dust Infrared Emission for Use in Estimation of Reddening and Cosmic Microwave Background Radiation Foregrounds, 500 (1998) 525–553. arXiv:astro-ph/9710327, doi:10.1086/305772.

\item K. I. Seon, J. Edelstein, E. Korpela, A. Witt, K.-W. Min, W. Han, J. Shinn, I.-J. Kim, J.-W. Park, Observation of the Far-ultraviolet Continuum Background with SPEAR/FIMS, 196 (2011) 15. arXiv:1006.4419, doi:10.1088/0067-0049/196/2/15.

\item K. I. Seon, A. Witt, I. J. Kim, J. H. Shinn, J. Edelstein, K. W. Min, W. Han, Comparison of the Diffuse H and FUV Continuum Backgrounds: On the Origins of the Diffuse H Background, 743 (2011) 188. arXiv:1111.2136, doi:10.1088/0004-637X/743/2/188.

\item E. T. Hamden, D. Schiminovich, M. Seibert, The Diffuse Galactic Far-ultraviolet Sky, 779 (2013) 180. arXiv:1311.0875, doi:10.1088/0004-637X/779/2/180.

\item D. Calzetti, R. C. Kennicutt, C. W. Engelbracht, C. Leitherer, B. T. Draine, L. Kewley, J. Moustakas, M. Sosey, D. A. Dale, K. D. Gordon, G. X. Helou, D. J. Hollenbach, L. Armus, G. Bendo, C. Bot, B. Buckalew, T. Jarrett, A. Li, M. Meyer, E. J. Murphy, M. Prescott, M. W. Regan, G. H. Rieke, H. Roussel, K. Sheth, J. D. T. Smith, M. D. Thornley, F. Walter, The Calibration of Mid-Infrared Star Formation Rate Indicators, 666 (2007) 870–895. arXiv:0705.3377, doi:10.1086/520082.

\item M. K. M. Prescott, R. C. Kennicutt, Jr., G. J. Bendo, B. A. Buckalew, D. Calzetti, C. W. Engelbracht, K. D. Gordon, D. J. Hollenbach, J. C. Lee, J. Moustakas, D. A. Dale, G. Helou, T. H. Jarrett, E. J. Murphy, J.-D. T. Smith, S. Akiyama, M. L. Sosey, The Incidence of Highly Obscured Star-forming Regions in SINGS Galaxies, 668 (2007) 182–202.
arXiv:0706.3501, doi:10.1086/521071.

\item S. C. Madden, Effects of massive star formation on the ISM of dwarf galaxies, 44 (2000) 249–256. arXiv:astro-ph/0002046, doi:10.1016/S1387-6473(00)00050-6.

\item E. Galliano, D. Alloin, E. Pantin, P. O. Lagage, O. Marco, Mid-infrared imaging of active galaxies. Active nuclei and embedded star clusters, 438 (2005) 803–820. arXiv:astro-ph/0504118, doi:10.1051/0004-6361:20053049.

\item A. N. Witt, B. Gold, F. S. Barnes, III, C. T. DeRoo, U. P. Vijh, G. J. Madsen, On the Origins of the High-latitude H Background, 724 (2010) 1551–1560. arXiv:1010.4361, doi:10.1088/0004-637X/724/2/1551.

\item J. Murthy, Modelling dust scattering in our Galaxy, 459 (2016) 1710–1720. arXiv:1601.00430, doi:10.1093/mnras/stw755.

\item K. Sano, K. Kawara, S. Matsuura, H. Kataza, T. Arai, Y. Matsuoka, Measurements of diffuse sky emission components in high galactic latitudes at 3.5 and 4.9 um using DIRBE AND WISE DATA. The Astrophysical Journal 818 (1) (2016) 72. URL http://stacks.iop.org/0004-637X/818/i=1/a=72

\end{enumerate}

\end{document}